\documentclass[twocolumn]{aastex62}
\usepackage{rotating}
\usepackage{amsmath}
\usepackage{cases}

\newcommand{\lgt}{$\log \left< T_e \right>$}

\shorttitle{The Application of the FBP Algorithm to SRT }
\shortauthors{Cho et al.}

\begin{document}

\title{The Application of the Filtered Backprojection Algorithm to Solar Rotational Tomography}

\author[0000-0001-7460-725X]{Kyuhyoun Cho}
\email{chokh@astro.snu.ac.kr}
\affil{Astronomy Program, Department of Physics and Astronomy, Seoul National University, Seoul 151-747, Republic of Korea}

\author[0000-0002-7073-868X]{Jongchul Chae}
\affil{Astronomy Program, Department of Physics and Astronomy, Seoul National University, Seoul 151-747, Republic of Korea}

\author[0000-0002-2106-9168]{Ryun-Young Kwon}
\affil{Korea Astronomy and Space Science Institute, Daejeon 34055, Republic of Korea}

\author[0000-0003-1859-0515]{Su-Chan Bong}
\affil{Korea Astronomy and Space Science Institute, Daejeon 34055, Republic of Korea}

\author[0000-0003-2161-9606]{Kyung-Suk Cho}
\affil{Korea Astronomy and Space Science Institute, Daejeon 34055, Republic of Korea}

\correspondingauthor{Jongchul Chae}
\email{jcchae@snu.ac.kr}



\begin{abstract}
 Solar rotational tomography (SRT) is an important method to reconstruct the physical parameters of the three-dimensional solar corona. Here we propose an approach to apply the filtered backprojection (FBP) algorithm to the SRT. The FBP algorithm is generally not suitable for SRT due to the several issues with solar extreme ultraviolet (EUV) observations, in particular a problem caused by missing data because of the unobserved back side of corona hidden behind the Sun. We developed a method to generate a modified sinogram which resolves the blocking problem. The modified sinogram is generated by combining the EUV data at two opposite sites observed by the Atmospheric Imaging Assembly (AIA) onboard the {\it Solar Dynamics Observatory} (SDO). We generated the modified sinogram for about one month in 2019 February and reconstructed the three-dimensional corona under the static state assumption. In order to obtain the physical parameters of the corona, we employed a DEM inversion method. We tested the performance of the FBP algorithm with the modified sinogram by comparing the reconstructed data with the observed EUV image, electron density models, previous studies of electron temperature, and an observed coronagraph image. The results illustrate that the FBP algorithm reasonably reconstructs the bright regions and the coronal holes, and can reproduce their physical parameters. The main advantage of the FBP algorithm is that it is easy to understand and computationally efficient. Thus, it enables us to easily probe the inhomogeneous coronal electron density and temperature distribution of the solar corona.
\end{abstract}



\section{Introduction} \label{sec:intro}

Tomography is a technique to infer the three-dimensional (3D) structure of an optically thin object by observing at various projection angles. Tomography can be categorized as either the filtered backprojection (FBP) algorithm or the iterative reconstruction \citep{Hsieh2015}. Each has pros and cons. The FBP algorithm is mainly used for radiology and has the advantage of easy and computational efficiency because it is based on an analytic solution; however, it requires a large number of observations at different projection angles. In contrast, iterative reconstruction can generate high quality data using only a few observations within known constraints. But it is computationally demanding for matrix calculations, at least 10 to 100 times more than that of the FBP algorithm \citep{Wang2016}. 

Solar rotational tomography (SRT) is the tomography of the solar corona making use of the solar rotation. The solar corona is optically thin and it can be observed at many different angles from a position as the Sun rotates. There are, however, several practical issues in the application of tomography to the solar corona directly \citep{Frazin2000}. One of the most critical issues is the problem caused by missing data due to large parts of the corona being hidden. A coronagraph blocks out all the light behind the occulter around the disk of the Sun. Even in the case of solar extreme ultraviolet (EUV) observations, the light from the backside is blocked by the Sun itself.  

Because of this issue, the iterative reconstruction is the most common approach. Even with a limited number of observational data, the iterative reconstruction has the advantage of finding the solution consistent with constraints or other a priori information, such as the information about the blocked area, by forward modeling. A number of relevant previous studies such as algebraic recontruction technique are reviewed by \citet{Frazin2000}. 

During the last two decades, the SRT using regularization methods was developed (See \citet{Frazin2000},  \citet{Aschwanden2011}, and references therein). It seeks the solution of the ill-posed problem by repeating the inversion for the best regularization parameters. One example is the robust, regularized, positive estimation method developed to determine the 3D electron density distribution from Large Angle and Spectrometric Coronagraph (LASCO) C2 data \citep{Frazin2002, Frazin2007}, Solar Terrestrial Relations Observatory (STEREO) COR1 \citep{Butala2010, Kramar2009}, and Extreme UltraViolet Imager (EUVI) data \citep{Kramar2014}. Several studies were conducted to estimate both electron density and temperature from the STEREO/EUVI data by combining the forward modeling techniques and differential emission measure (DEM) inversion \citep{Frazin2009, Vasquez2009, Vasquez2011}. Recently, a new method of determining coronal electron density using spherical harmonics and mass flux conservation for a spherically expanding corona was proposed \citep{Morgan2015, Morgan2019a}. 

Nevertheless, the FBP algorithm is still attractive because it offers greater efficiency. It can provide a synoptic view for the solar corona if the aforementioned issue is solved. There have been a few attempts to utilize the FBP algorithm in highly restricted cases so far. For example, polar plumes were studied using the FBP algorithm \citep{dePatoul2013}. It was possible because the polar plumes are located on the polar regions that are not affected by blocking. \citet{Morgan2009} exploited the FBP algorithm to reconstruct coronal structures from LASCO/C2 data. The main idea is that the coronal structures may be assumed to be radial beyond 3 R$_\odot$. Assuming that the radial structures extend to the center of the Sun, the FBP algorithm is applicable.  

In this paper, we present a new approach to reconstruct the 3D structure of the solar corona using the FBP algorithm in the EUV regime. We developed a method to generate a modified sinogram. The modified sinogram allows us to reconstruct the solar corona from the solar EUV observations with the unobservable backside being properly taken into account. By combining the modified sinogram using data taken by the Atmospheric Imaging Assembly (AIA; \citealt{Lemen2012}) onboard the {\it Solar Dynamics Observatory} (SDO), and the assumption of the static state corona, we propose to resolve the blocking backside issue of the SRT. We test the performance of our results by comparing it with observations, models, and previous studies. Finally, we will discuss its limitations and applications. 

\section{Method} \label{sec:method}

\begin{figure}
\includegraphics[width=0.5 \textwidth]{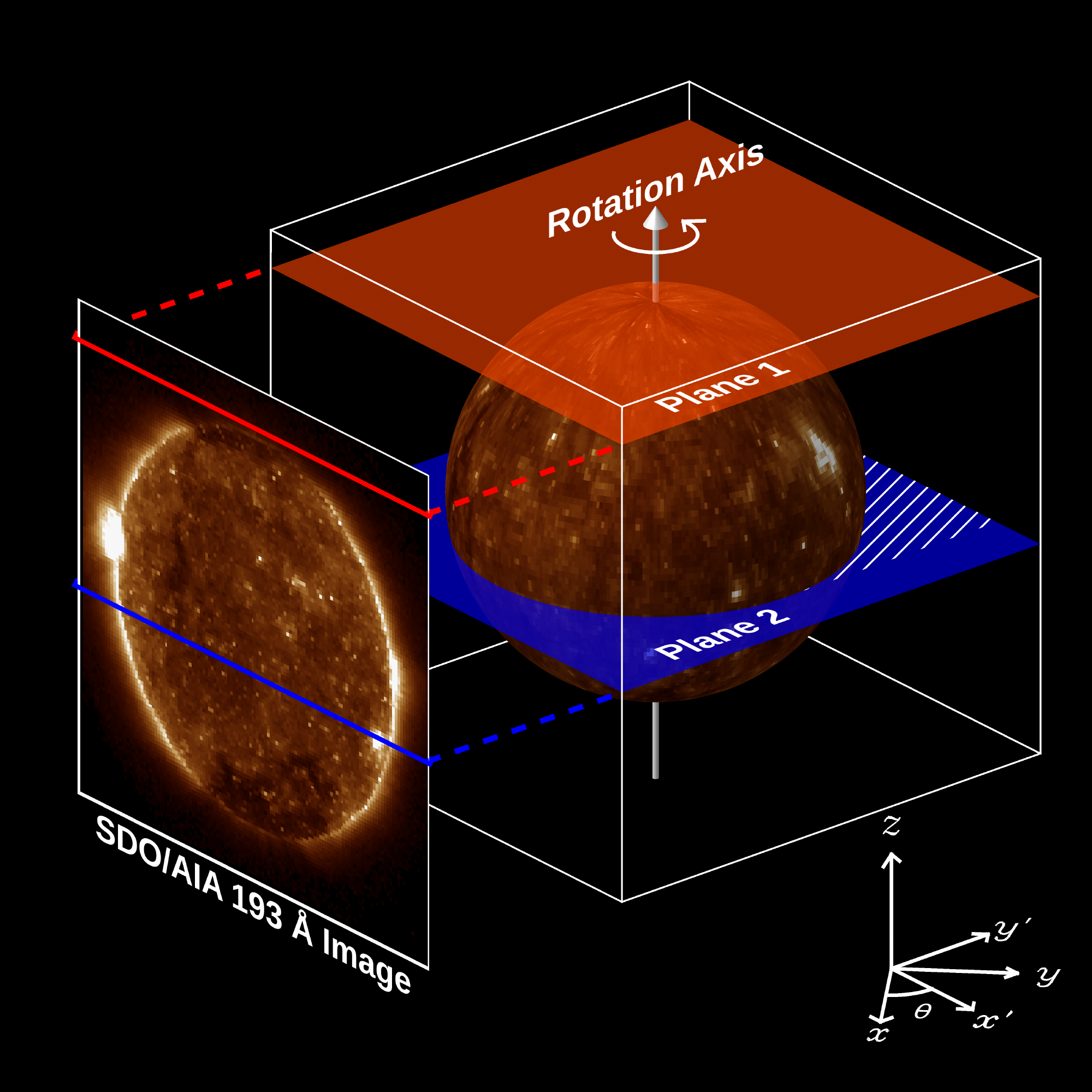}
\caption{An illustration of the SRT.  The 3D view of the solar corona and projected 2D SDO/AIA image are presented. Plane 1 (red) and Plane 2 (blue) represent latitudinal planes over the polar region and intersecting with the Sun, respectively. The white hatched area in Plane 2 indicates the blocking area by the Sun.}
\label{fig1}
\end{figure}

The reconstruction of a 3D volume can be understood as a series of two-dimensional (2D) reconstructions. In the case of the SRT, we obtain the 3D information of the solar corona by ignoring the inclination angle of the observer and by stacking the reconstructions of each latitudinal plane, which is perpendicular to the solar rotation axis (Figure \ref{fig1}). Thus, the problem changes into how to successfully reconstruct the 2D distribution of the physical quantities from observations. 

\subsection{Conventional FBP Algorithm}
We briefly review the formulation of the two-dimesional FBP algorithm here. The details are described in the work of  \citet{Hsieh2015}. The FBP algorithm is based on the Fourier slice theorem that one projection of the object can fill a line in the 2D Fourier domain. One can obtain a sinogram $p(x', \theta)$, which is a parallel projection at angle $\theta$ of a 2D object $f(x, y)$
\begin{equation}\label{projection}
p(x', \theta) = \int_{-\infty}^{\infty} {f(x, y)} dy'  
\end{equation}
where $(x, y)$ and $(x', y')$ are the coordinates in the object and observer frames, respectively, and $\theta$ is the angle between the two coordinates. The relation between two coordinate systems can be expressed as:
\begin{eqnarray}
x' & = & x \cos \theta + y \sin \theta \\
y' & = & -x \sin \theta + y \cos \theta .
\end{eqnarray}
In the observational data, where solar north is the image vertical, the variable $x'$ corresponds to the index of the row array in the CCD as illustrated in Figure \ref{fig1}, and $\theta$ varies with observing time. Thus, a sinogram consists of a bundle of the fixed row array data with different observing times. The sinogram may have the values of either intensity or emission measure (EM).

If the Fourier transform of the sinogram $p(x',\theta)$ over the variable $x'$ is denoted by $P(\omega, \theta)$, we acquire the following relation 
\begin{eqnarray}\label{Fourier pair}
P(\omega, \theta) & = & \int_{-\infty}^{\infty}{p(x', \theta) e^{-i 2 \pi \omega x'} dx'}  \\ 
& = & \int_{-\infty}^{\infty} \int_{-\infty}^{\infty} {f(x, y) e^{-i 2 \pi \omega (x \cos \theta + y \sin \theta)} dx' dy'} \\
& = & F(\omega  \cos \theta, \omega \sin \theta)  
\end{eqnarray}
which states that $P(\omega, \theta)$ are the values of $F(k_x, k_y)$, the 2D Fourier transform of $f(x, y)$, on the straight line defined by $k_x = \omega \cos \theta$ and $k_y= \omega \sin \theta$. If the 2D Fourier domain can be filled with projections at many different angles, we can reconstruct $f(x, y)$ by the inverse Fourier transform in the polar coordinate
\begin{equation}
f(x, y) =  \int_{0}^{2\pi} \int_{0}^{\infty}{P(\omega, \theta)   \omega e^{i 2 \pi \omega (x \cos \theta + y \sin \theta)} d\omega}{d\theta}.
\end{equation}
The observed data from two opposing viewpoints have identical but inverted information 
\begin{equation}
p(x', \theta+\pi) = p(-x', \theta).
\end{equation}
Thus data taken over half a rotation are sufficient for reconstruction
\begin{equation}\label{FBP algorithm}
f(x, y) =  \int_{0}^{\pi} \int_{-\infty}^{\infty}{P(\omega, \theta) \left | \omega \right | e^{i 2 \pi \omega (x \cos \theta + y \sin \theta)} d\omega}{d\theta}.
\end{equation}

To make a better reconstruction, we need to fill the 2D Fourier domain with dense observations. This is the reason why the FBP algorithm requires lots of observational data at different angles. According to the Nyquist sampling theory, it is known that the required minimal number of observations $n_\theta$ for the FBP algorithm is 
\begin{equation}
n_\theta \ge \frac{\pi}{2}n_R.
\end{equation}
where $n_R$ is the reconstructed data size. For the SRT, considering the equatorial rotation period of 25 days, the required time cadence $\Delta t$ is 
\begin{equation}
\Delta t \le 2.3 \times 10^4 / n_R \, \textrm{min}.
\end{equation}
For example, if the size of the reconstructed data is 128 pixels, the minimal time cadence is about 180 minutes. But the shorter time cadence ensures better quality because the interpolation is used to fill the 2D Fourier domain.

\begin{figure*}
\plotone{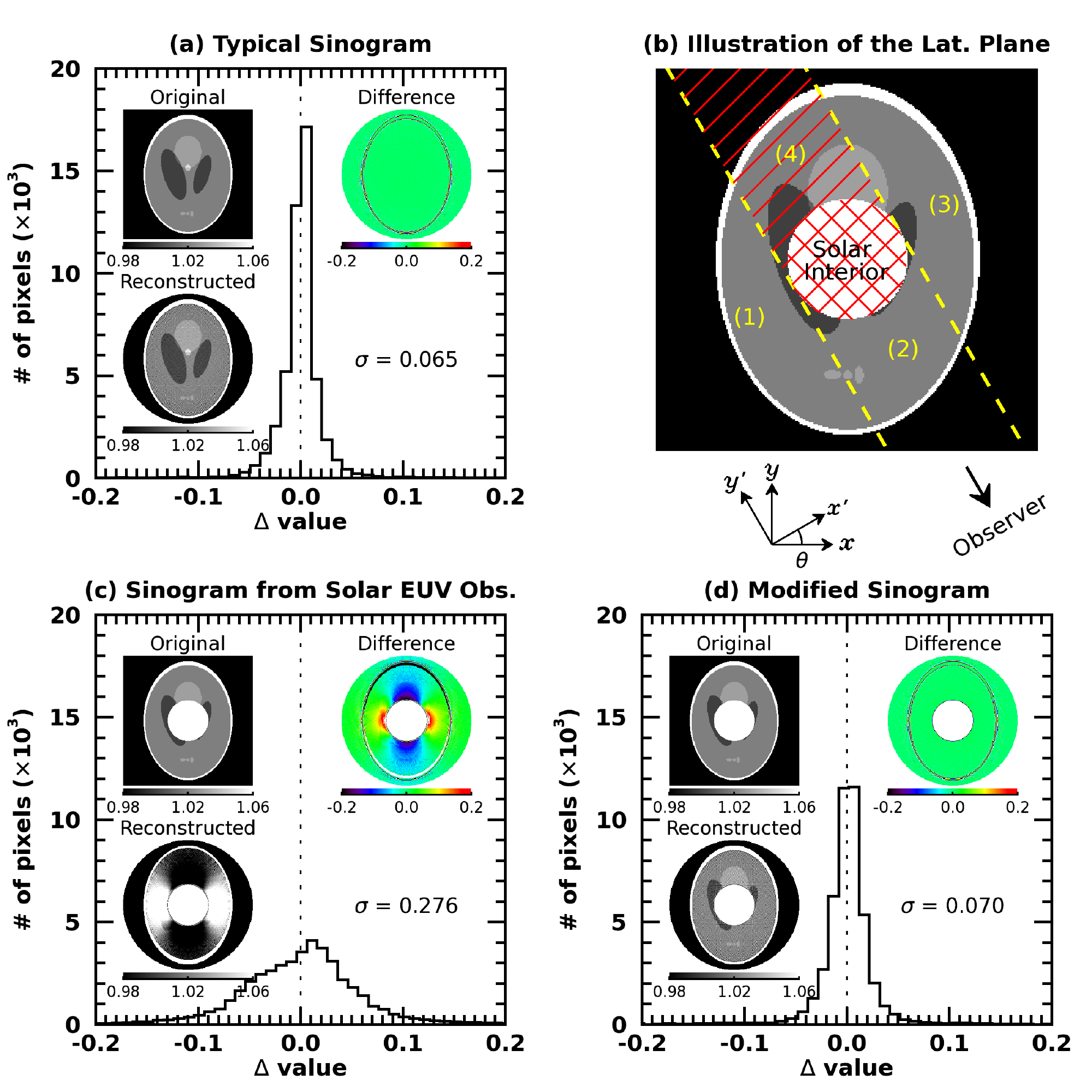}
\caption{(a) Histogram of the difference image using the FBP algorithm with an optically thin object. The difference image is found by subtracting the reconstructed image from the original image.  (b) An illustration of a latitudinal plane for the solar EUV observation which is similar to plane 2 in Figure \ref{fig1}. The direction of an observer is shown as the black arrow. (1) to (4) represent classified areas by observability. (4) shown as red hatched area indicates the backside of the solar corona, which is the same with the hatched area on plane 2 in Figure \ref{fig1}. The red crosshatched area near the image center indicates the solar interior. Two coordinate systems are presented below the image. (c) Histogram of the difference image using the FBP algorithm with the sinogram made by solar EUV observations. (d) Histogram of the difference image using the FBP algorithm with the modified sinogram. The bin size of the histograms is 0.01. The original, reconstructed, and difference images are included in each histogram plot. The corresponding standard deviation $\sigma$ is shown. }
\label{fig2}
\end{figure*}

We examined the performance of the FBP algorithm using the Shepp-Logan phantom \citep{Shepp1974} with a size of 256 $\times$ 256 pixels as a test image. The area in the Shepp-Logan phantom is divided by 10 ellipses and has levels ranging from 0 to 2, especially from 1 to 1.04 near the center. We made one-dimensional pseudo observation data which are the integration of the original image at different angles with every 0.5 degrees. The sinogram is produced by collecting these pseudo data. Then we apply the sinogram to the FBP algorithm for generating the reconstructed image. 

Figure \ref{fig2}a indicates the reconstructed result of the FBP algorithm using a test image. The difference image between the original and the reconstructed image shows that most pixels are close to zero value except at rapidly changing boundaries. The histogram of the difference image shows that most of the pixels have values near zero with a standard deviation of 0.065. This shows that the FBP algorithm successfully reconstructed the original image.

\subsection{SRT with Modified Sinogram}
Solar EUV observations are different from the typical condition for the FBP algorithm. As explained in the section \ref{sec:intro}, the FBP algorithm is valid for an optically thin object. This condition well matches with the solar EUV observations over the polar region, where there is no missing data due to blockage by the Sun. It is well illustrated by plane 1 in Figure \ref{fig1}. But the situation differs in the non-polar region. When reconstructing a latitudinal plane including the Sun shown as plane 2 in Figure \ref{fig1}, the solar EUV observations are unavoidably affected by the Sun itself as an occulter. The detailed situation is illustrated in Figure \ref{fig2}b. Basically, we cannot obtain the information about the solar interior. Additionally, the backside (area 4) is screened by the Sun itself. In this state, the formulation of the sinogram from the solar EUV observations $p_{\odot}(x', \theta, z)$ is changed

\begin{eqnarray}\label{new_env}
&&p_{\odot}(x', \theta, z)  =  \int_{-\infty}^{s_{\textrm{eff}}}{f(x, y, z) dy'} \\
&&s_{\textrm{eff}}  =  
\begin{cases}
-\sqrt{R_\odot^2 -z^2-x'^2}, &  z < R_\odot \textrm{ and } \left | x' \right | < \sqrt{R_\odot^2-z^2}   \\
\infty, & \textrm{otherwise}
\end{cases}
\end{eqnarray}
where $R_\odot$ is the solar radius and the $z$-axis is parallel to the solar rotation axis. The interval of the integration represented by $s_\textrm{eff}$ is quite different from that of the optically thin case. 

It is impossible to obtain information about the solar interior at any angle, therefore it is regarded as a cavity. In contrast, area 4 cannot be treated as a cavity. It contributes to the other line of sight integrations at different observation angles. 

Figure \ref{fig2}c represents the result of the FBP algorithm using the sinogram from the Sheep-Logan test data with an induced cavity and missing backside data, similar to solar EUV observations. The sinogram $p_{\odot}(x', \theta)$ is made by the line of sight integration without the solar interior region and the backside at different angles as formulated in Equation \ref{new_env}. Compared with the result of the optically thin case, the FBP algorithm fails to precisely reconstruct the solar EUV observation; The difference image clearly shows systematic artifacts. Obviously, the histogram of the difference image also has a large deviation from the zero value. The standard deviation is more than 4 times larger than that of the complete case. 

Information from the missing backside region can be obtained by the observation at the opposite site. By combining data from the two opposing sides we are able to generate a complete sinogram $p(x', \theta, z)$ for the SRT
\begin{eqnarray}\label{mod_sino}
&&p(x', \theta, z) \\
&& =  \left \{ \begin{array}{ll}
p_{\odot}(x', \theta, z)+p_{\odot}(-x', \theta+\pi, z),  \nonumber \\
 \qquad \qquad \qquad z < R_\odot \textrm{ and } \left | x' \right | < \sqrt{R_\odot^2-z^2} \\
p_{\odot}(x', \theta, z),  \qquad  \textrm{otherwise}
\end{array} \right .
\end{eqnarray}

Each term in the first case of Equation \ref{mod_sino} indicates regions 2 and 4, respectively, and the second case indicates both areas 1 and 3 in Figure \ref{fig2}b. It enables us to generate the whole sinogram applicable to the FBP algorithm. We refer to this as the modified sinogram. Now we use the modified sinogram $p(x', \theta, z)$ defined by Equation \ref{mod_sino}, and execute the identical process presented in Equation  \ref{Fourier pair} and \ref{FBP algorithm} for every latitudinal planes. 

The result of the FBP algorithm with the modified sinogram is presented in Figure \ref{fig2}d. The difference image and its histogram demonstrate that the reconstruction is adequately performed. The difference image indicates that the reconstructed image is strikingly similar to the original image. The standard deviation of the difference image histogram is 0.070, which is comparable to the result of the FBP algorithm with the optically thin object.

\subsection{Validation of Parallel Beam Assumption} \label{subsec:fanbeam}

\begin{figure*}
\plotone{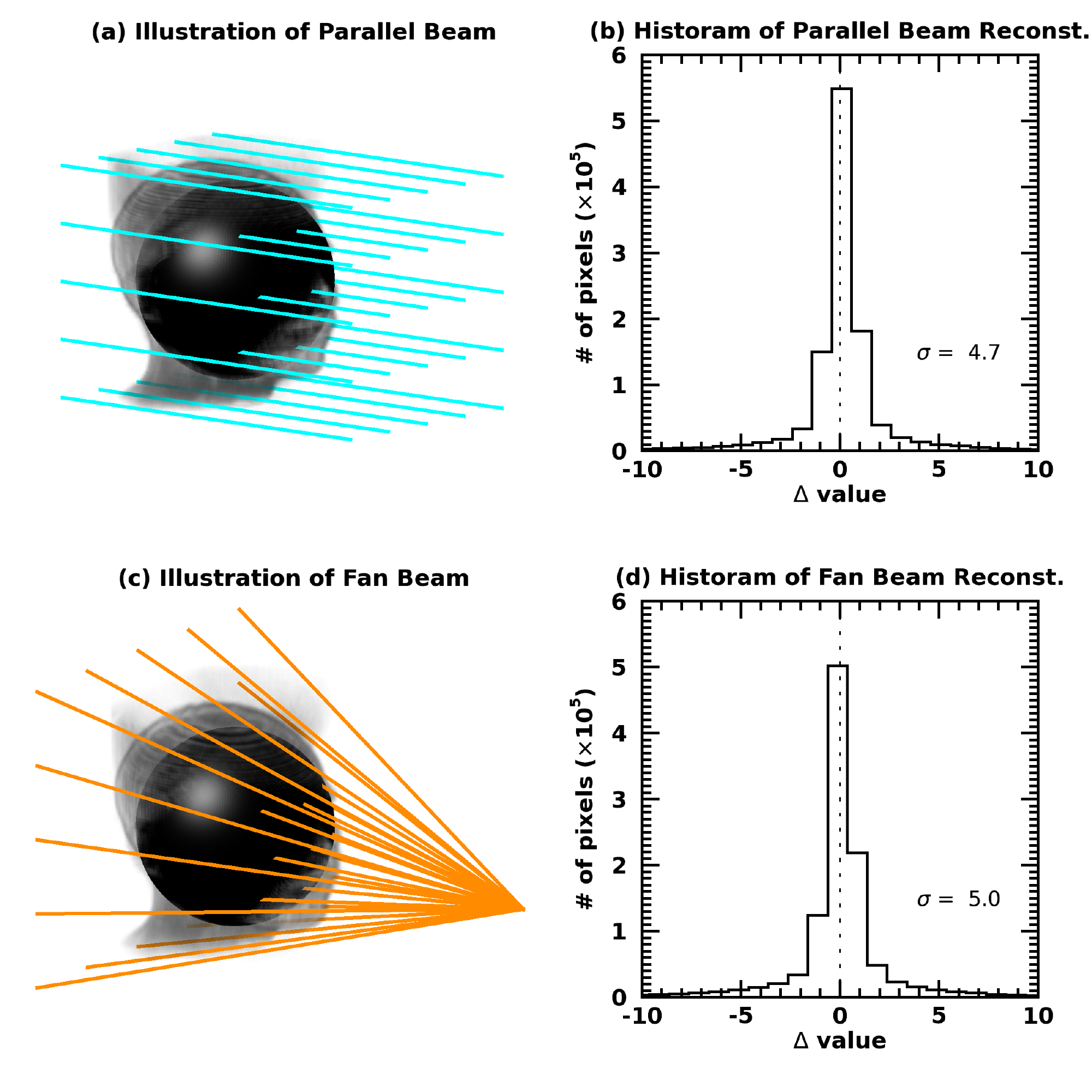}
\caption{(a) An illustration of the parallel beam projection. The black sphere and head-shaped gray volume indicate the solar interior and a coronal structure. The sky blue lines represent ray paths of the parallel beam projection. The ray cannot pass through the black sphere. (b) Histogram of the difference cube using FBP algorithm with a modified sinogram in the parallel beam projection. The difference cube is found by subtracting the each reconstructed voxel value from the original voxel value. (c) An illustration of the fan-beam projection. The orange lines represent ray paths of the fan-beam projection. Note that fan-beam angles are exaggerated about 100 times. Others are the same as (a). (d) Histogram of the difference cube using FBP algorithm with a modified sinogram in the fan-beam projection. }
\label{fig3}
\end{figure*}

We assumed the parallel beam projection for the SRT. In reality, however, ray paths are not parallel to each other because the Sun is much larger than a detector size. Thus it must be fan-beam projection. Figure \ref{fig3}a and \ref{fig3}c illustrate each projection. Reconstruction of 3D volume under the fan-beam projection is different from the parallel case \citep{Kak2001}. Moreover, we cannot generate a complete modified sinogram under the fan-beam projection because the area behind the Sun cannot be supplemented by observed data at the opposite side. So we have to assume the parallel beam projection for SRT. 

We tested the error caused by the parallel beam assumption. We made a 3D test cube using human head data built-in IDL (see Figure \ref{fig3}a and \ref{fig3}c). Its size is 128$^3$, which corresponds to 32 $\times$ 32 binning SDO/AIA data, and each voxel has a value between 0 and 255. The solar interior represented as black spheres was set up at the center of the test cube with a radius of 51 pixels. We generated pseudo observation data in both parallel and fan-beam projection. The observing rays cannot pass through the interior. We generated modified sinogram using Equation \ref{mod_sino} under the parallel beam assumption for both, and we reconstructed 3D cubes. Then we analyzed the difference cubes similar to Figure \ref{fig2}.

Our result clearly shows that the parallel beam assumption is valid for the SRT (Figure \ref{fig3}b and \ref{fig3}d). The standard deviation of the difference cube made by the parallel beam projection is 4.7. In a similar way, the standard deviation made by fan-beam projection is 5.0, which does not differ much from the above value. It implies that the distance of the Earth from the Sun is far enough to assume the parallel beam projection; in fact, the fan-beam deviates less than one voxel from the parallel beam in our test cube. 

\subsection{Axial Tilt Effect}

\begin{figure*}
\plotone{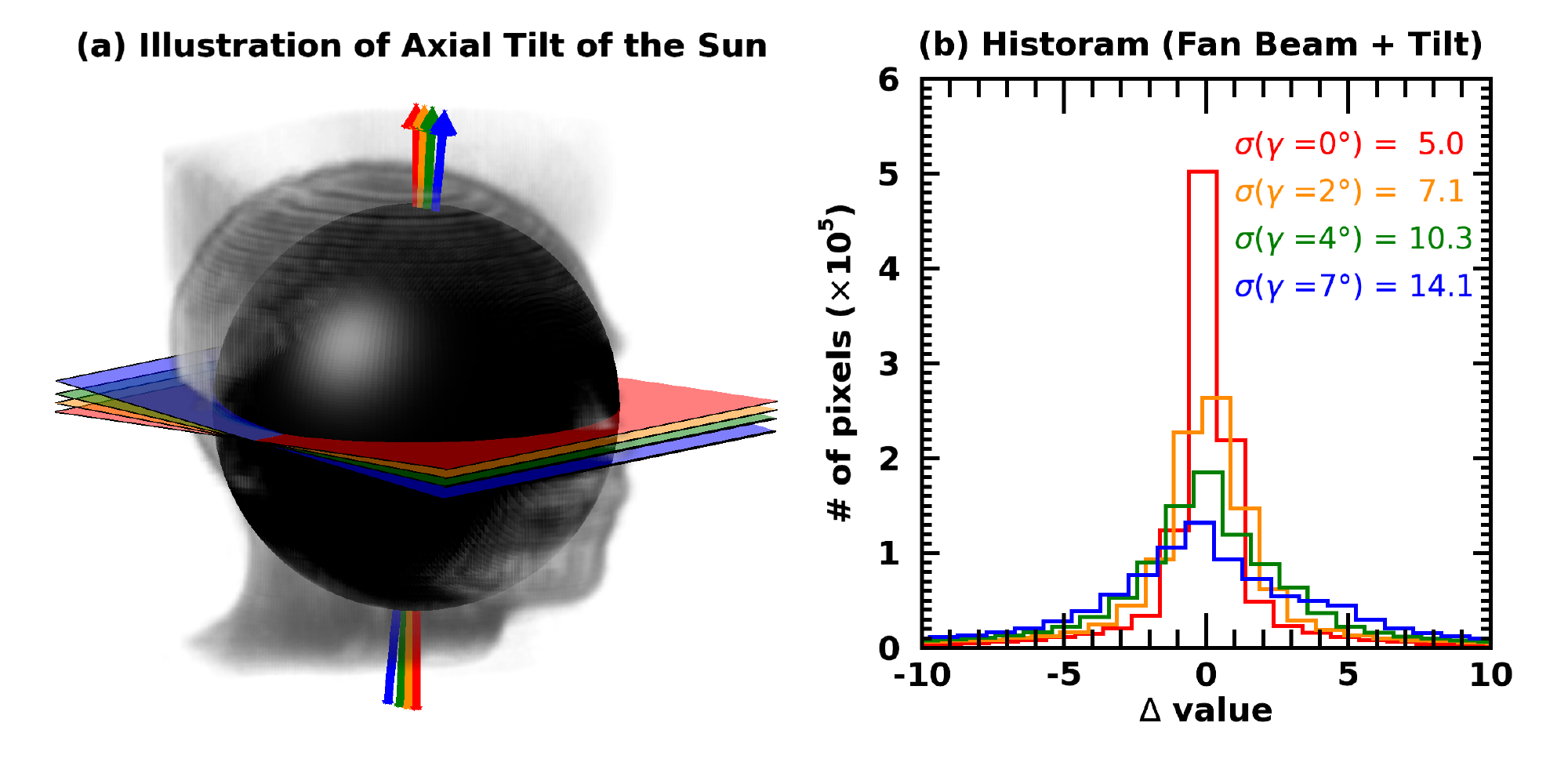}
\caption{(a) An illustration of the solar axial tilt. The arrows and planes indicate rotation axes and equatorial planes, respectively. Others are the same as Figure \ref{fig3}a. (b) Histogram of the difference cube using FBP algorithm with modified sinograms. Each color represents different tilt angle $\gamma$ from 0\degr\ to 7\degr\ in both panels. }
\label{fig4}
\end{figure*}

Another important assumption is that the rotational axis is perpendicular to the ecliptic plane. As a matter of fact, the axial tilt angle of the Sun varies between $\pm$7 degrees. It has a maximum and minimum value in September and February and is close to zero in June and December. The obtained modified sinogram deviates from real values as the tilt angle increases, so the error increases. We measured the error caused by the axial tilt angle. The pseudo observation data were generated considering the fan-beam projection when the tilt angle is 0\degr, 2\degr, 4\degr, and 7\degr\  (See Figure \ref{fig4}a). Then we analyzed the standard deviation of the difference cube as in Section \ref{subsec:fanbeam}. 

Figure \ref{fig4}b shows the results of analyses. The standard deviation increases with the axial tilt angle. The maximum standard deviation occurred when the axial tilt angle is 7\degr , and is about three times that of the zero tilt angle case. Therefore, data observed near June or December with almost zero tilt angle guarantees better reconstruction quality. However, we confirm that the major source of the error is not the axial tilt effect, but the temporal evolution of the solar coronal structures. We will discuss that in Section \ref{sec:diss}. 

\section{Data and Analysis} \label{sec:data}
SDO/AIA data is used for the reconstruction of the coronal structures. We supposed that the solar corona is in a static state during one solar rotation. We took a period without sunspots, 2019 February 15 $\pm$ 17 days to minimize the error from the temporal evolution of the solar corona. The data taken through six EUV channels (94 \AA, 131 \AA, 171 \AA, 193 \AA, 211 \AA, 335 \AA) were collected every 30 minutes. After manually removing inappropriate data taken during eclipses or spacecraft maneuvers, we executed {\tt\string aia\_prep.pro} on 1637 observations $\times$ 6 channel data to obtain the AIA level 1.5 data that was aligned and normalized by their exposure time. The 4k$\times$4k data were rebinned to 128$\times$128 pixels to reduce the computing time. The result is six 3D datacubes of intensity values in units of DN/s, $I_{w}(x', t, z)$, where $t$ is the observing time and $w$ is AIA EUV channel. 
\begin{figure*}
\plotone{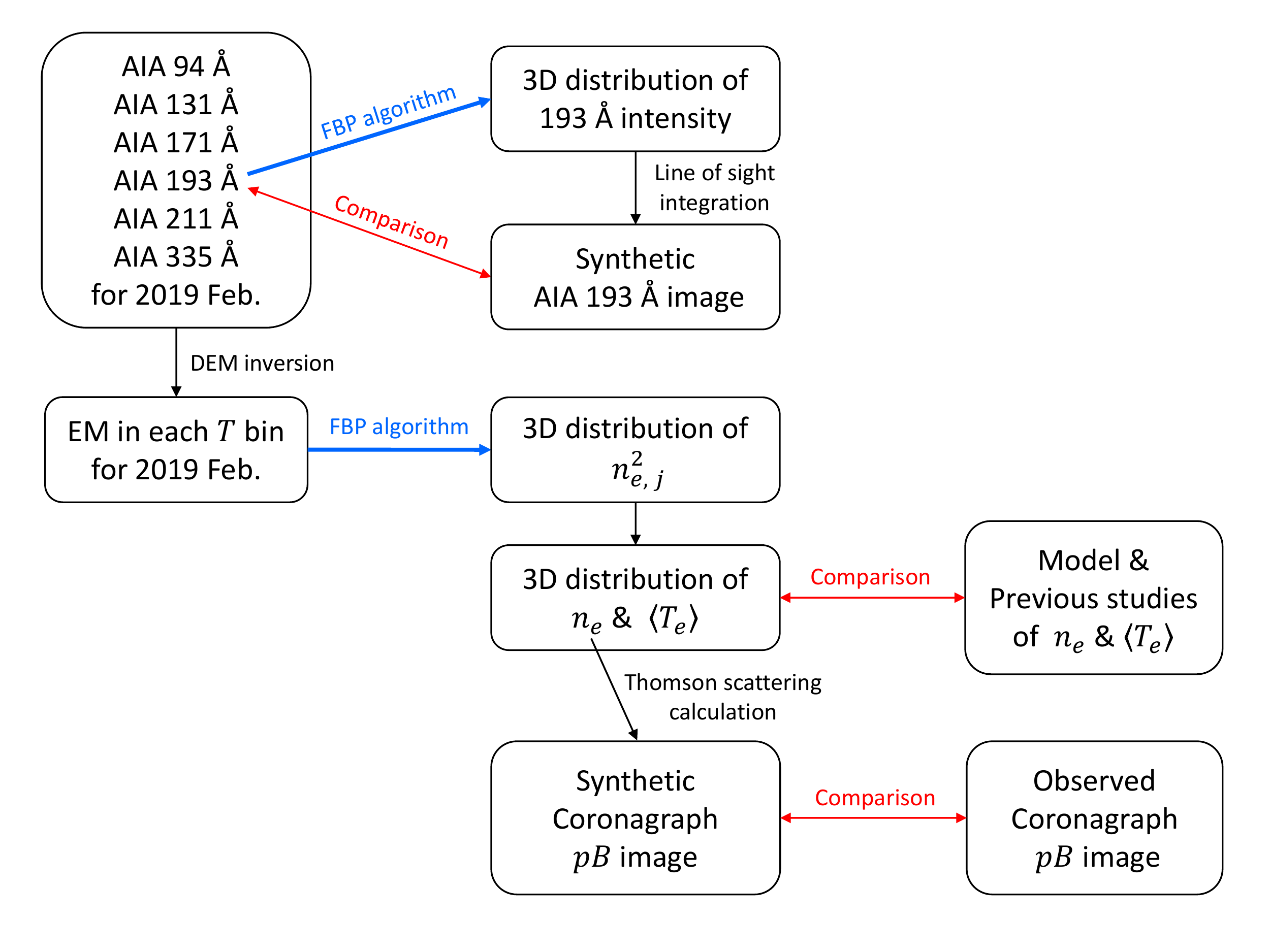}
\caption{A flowchart for the test of the FBP algorithm using SDO/AIA data. }
\label{fig5}
\end{figure*}

We tested the FBP algorithm in three ways. These are illustrated in Figure \ref{fig5}, and each result constitutes a subsection of Section \ref{sec:res}. First, we used the synthetic 193 \AA\ intensity images. We made the modified sinogram from $I_{193}(x', t, z)$. We assumed the tilt angle of the SDO orbit from the ecliptic plane to be negligible. The differential rotation \citep{Snodgrass1990} was considered for each latitudinal plane with a reference time of 2019 February 15. Thus the observing time $t$ was converted to the observing angle $\theta$, and the reference time was set to zero observing angle. The FBP algorithm reconstructed the 3D distribution of 193 \AA\ intensity. The source code of the FBP algorithm was written in IDL. It took less than 2 minutes to reconstruct each cube on a 2.93 GHz Intel Xeon CPU. By using Equation \ref{new_env}, we obtained the synthetic 193 \AA\ images at different angles. These images were compared to the observed SDO/AIA 193 \AA\ images that have been used as the input for the reconstruction. 

Second, we determined the distribution of total electron density $n_e$ and that of electron temperature $T_e$ in the solar corona. To obtain the physical quantities, we exploited the method for DEM inversions developed by \citet{Cheung2015} to the AIA level 1.5 data. We used the default setting $\sigma = [0, 0.1, 0.2, 0.6]$ as the Gaussian basis function matrix, and set 10 temperature bins from $\log T=5.5$ to $\log T=6.5$ with a bin size of $\Delta \log T = 0.1$ because high noise in the DEM map occurs at temperature higher than $\log T = 6.5$. The output of the DEM inversion is the set of EM$_{j}(x', t, z)$ in the $j$-th bin of the $\log T$.

The same FBP algorithm process was applied to EM$_{j}(x', t, z)$.  According to Equation \ref{new_env} and the definition of EM
\begin{eqnarray}\label{EM ne}
\textrm{EM}_{j}(x', \theta, z) & = & \int_{-\infty}^{s_{\textrm{eff}}}{n_{e, j}^2(x, y, z) dy'},
\end{eqnarray}
each voxel has information about $n_{e, j}^2$, which corresponds to the emissivity in the given temperature bin. Some voxels may have negative values of $n_{e,j}^2$ because of mathematical error in the reconstruction and because of the lack of positivity constraint \citep{Frazin2009}. The negative values are unphysical, so they were replaced with zero for further analysis. The total electron density and the electron temperature for each voxel were calculated using the formulae:
\begin{eqnarray}
n_e(x, y, z) = \sqrt{ \sum_{j} n_{e, j}^2(x, y, z)} \label{eq_ne} \\
T_e (x, y, z) = \frac{1}{n_e^2} \sum_{j} n_{e, j}^2(x, y, z) T_{j} . \label{eq_te}
\end{eqnarray}
Note that $T_e$ represents the temperature weighted by the square of electron density and is undefined at the zero-density voxels. Similarly, a mean electron temperature of a defined region of interest $\left< T_e \right >$ was calculated by 
\begin{equation}
\left< T_e\right> = \frac{\sum_{j} (\int_{V} n_{e, j}^2 dV') T_j }{\sum_{j}\int_{V} n_{e, j}^2 dV'}, 
\end{equation}
where $V$ is the volume of the region of interest. The calculated physical parameters were compared with the previously reported models of electron density and electron temperature. 

Third, we made a synthetic coronagraph image. From the total electron density distribution we calculated radially polarized intensity $I_r$, and tangentially polarized intensity $I_t$ using the Thomson scattering calculation \citep{Cho2016} near the 7500 \AA, and generated the polarized brightness ($pB$) image 
\begin{equation}
pB = I_t-I_r.
\end{equation}
We compared the synthetic coronagraph $pB$ image with an observed $pB$ image taken by the K-Corongraph (K-Cor) of the COronal Solar Magnetism Observatory (COSMO)\footnote{\href{https://www2.hao.ucar.edu/mlso/instruments/mlso-kcor-coronagraph}{https://www2.hao.ucar.edu/mlso/instruments/mlso-kcor-coronagraph}} at the reference time. 
\section{Results and Performance Tests} \label{sec:res}
\subsection{Synthetic 193 \AA\ Image} \label{sec:res_1}
\begin{figure*}
\plotone{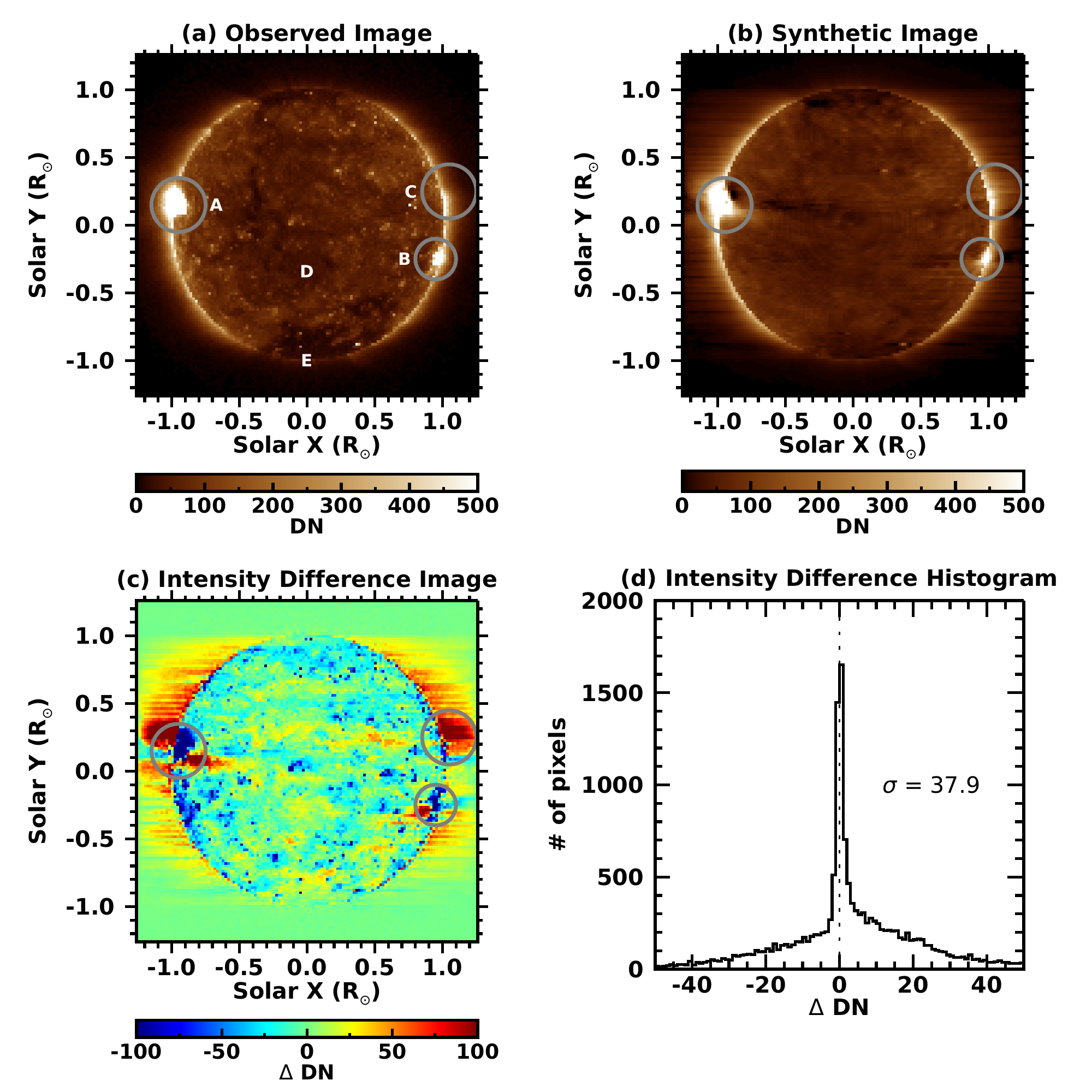}
\caption{(a) SDO/AIA 193 \AA\ image at 2019 Feb 15 00:00:05 UT.  (b) Synthetic 193 \AA\ image using the FBP algorithm. (c) Difference image between the observed image and the synthetic image. Region A, B, and C circled in (a)-(c) represent areas showing the discrepancy between two images. (d) Histogram of the difference image. The bin size of the histogram is 1. The corresponding standard deviation is specified together. The animation shows the comparison between observed images (left) and synthetic images (right) in the different angle views. D and E in (a) indicate regions of interest for detailed analysis in Figure \ref{fig9}.}
\label{fig6}
\end{figure*}

We found that the reconstructed 193 \AA\ image is very similar to the observed image (Figure \ref{fig6}a, \ref{fig6}b).  A few bright areas (e.g. region A and B) and coronal holes located in the north and south pole were successfully reconstructed in the same regions as in the observed image, even though their shapes slightly differ from those in the observed image. We found that long-lasting objects were successfully reconstructed, whereas most of the small short-lived bright points could not be properly reconstructed. 

The difference in intensity between the two images is noticeable in the three regions marked by A, B, and C (Figure \ref{fig6}c). Region A and B correspond to bright regions in the observed image, and region C, to another bright region located off the limb (see the associated animation). They have elongated features mostly along the longitudinal direction. They are artifacts resulting from the temporal evolution of the coronal loops. Note that the successful performance of the FBP algorithm requires no temporal variation of features during the observation. The north and south coronal holes were fairly well reconstructed because they stayed stable for more than half the solar rotation period. 

The histogram of intensity difference indicates that most of the difference values are around zero with a standard deviation of 37.9 (Figure \ref{fig6}d). This histogram is different from the one in Figure \ref{fig2}d in that it consists of two components: the narrow component that is commonly seen in both the histograms and the broad component that is seen only in this histogram. This broad component corresponds to the artifacts as mentioned above. 

The animation associated with Figure \ref{fig6} displays the series of synthetic 193 \AA\ images seen at different angles and the corresponding observed SDO/AIA 193 \AA\ images. The latter half of the animation displays the x-axis rotation view, so we can look over the polar region of the solar corona. The animation indicates that the two sets of images are quite similar to each other, especially in the morphology of regions A, B, C and both coronal holes. It is  interesting that the FBP algorithm partly reflects the time evolution of the coronal structures in the reconstruction. For example, region B did not appear until 2019 February 12 in the observation images, and the corresponding synthetic  images closely mimic the feature of region B. It implies that the FBP algorithm reflected the time evolution of the corona in the 3D structure. The reconstructed data carried some information on the temporal evolution because the reconstruction was done based on the static corona assumption, which results in a similar effect of time average.

\subsection{3D distribution of $n_e$ \& $T_e$ }
\begin{figure*}
\plotone{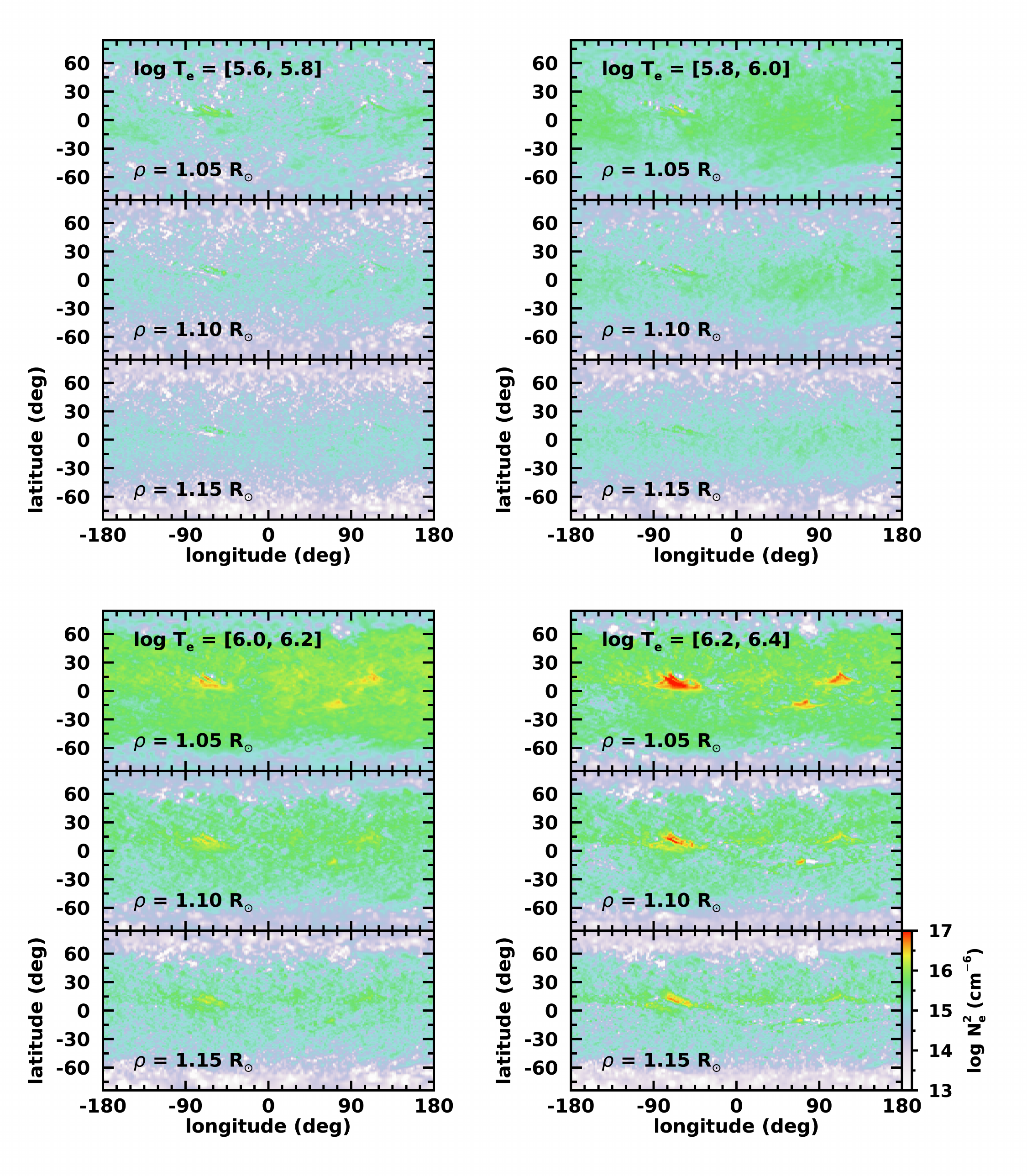}
\caption{Square of electron number density maps at different heights within the given temperature range. Each panel indicates different temperature from log $T_e$ = 5.6 to 6.4 at an interval of $\Delta$log $T_e$ = 0.2. Each row in the panels represents the height of 1.05, 1.10, 1.15 R$_\odot$, respectively. Zero longitude indicates the Sun's central meridian at the reference time.}
\label{fig7}
\end{figure*}

\begin{figure*}
\plotone{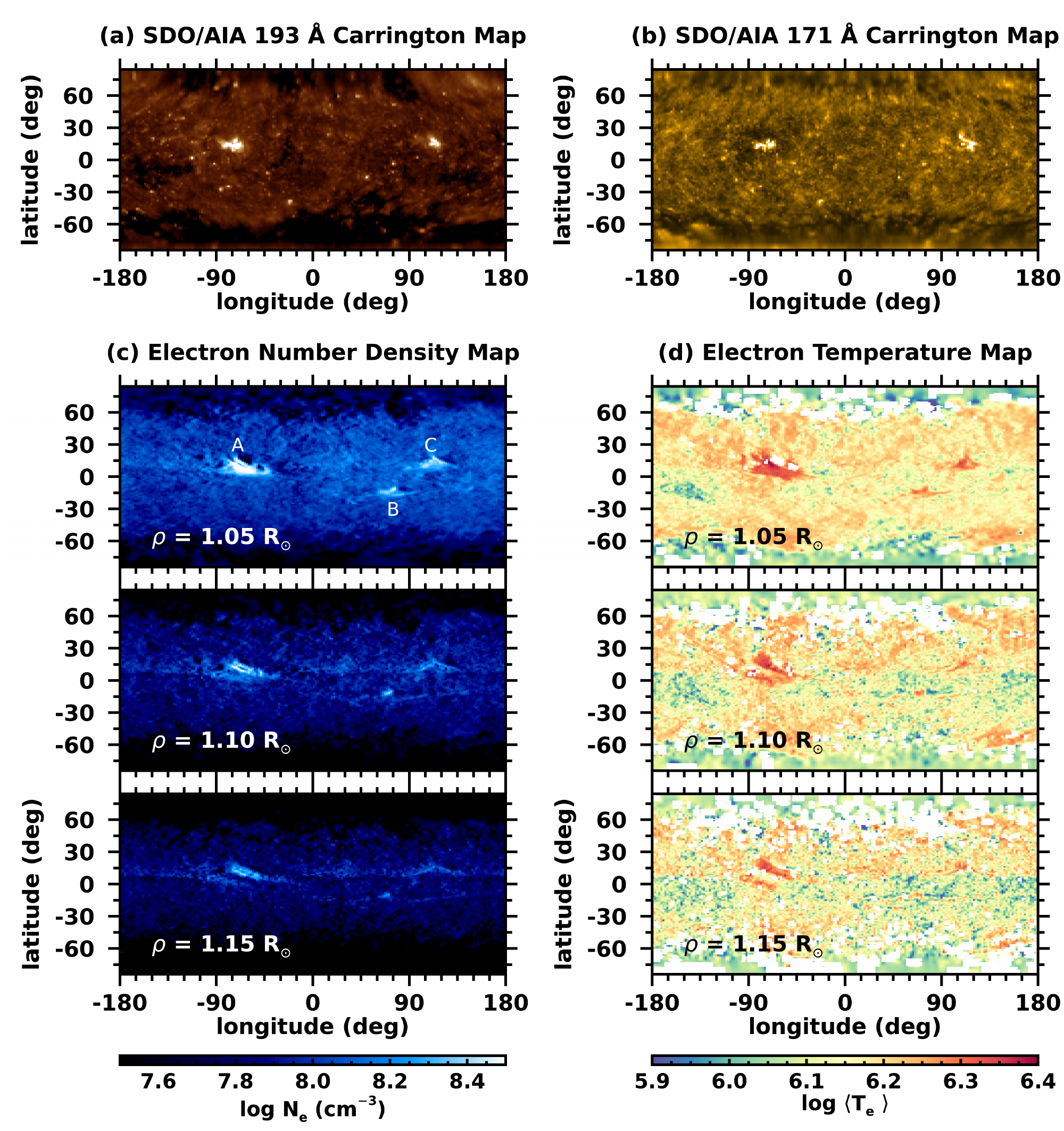}
\caption{(a) SDO/AIA 193 \AA\ Carrington map. (b) SDO/AIA 171 \AA\ Carrington Map. Zero longitude indicates 2019. 2. 15 00:00:05 UT. (c) Electron number density maps at different heights. The regions marked by A, B, and C indicate identical areas with Figure \ref{fig6}a. (d) Electron temperature maps at different heights. The blank regions indicate the undefined temperature region. Each raw represents the height of 1.05, 1.10, 1.15 R$_\odot$, respectively. Zero longitude indicates the Sun's central meridian at the reference time. The animation shows cross sections of the reconstructed coronal electron density cube as the distance from the solar center changes.}
\label{fig8}
\end{figure*}

Figure \ref{fig7} shows square of electron density distributions at different heights and in different temperature ranges. From this results, Equation \ref{eq_ne}, and \ref{eq_te}, we can obtain 3D electron density and temperature distributions (Figure \ref{fig8}). Figure \ref{fig8}c shows the map of the calculated electron density at each height. The associated animation gives the information about the 3D electron density distribution. As expected, we found that the regions of lower altitude and lower latitude have higher density in both the maps and the animation. The bright regions in SDO/AIA images (Figure \ref{fig8}a, \ref{fig8}b) well match the highest density region marked by A, B, and C.

Figure \ref{fig8}d shows the map of electron temperature at each height. The temperature map is similar to the electron density map in that the higher altitude region generally has a lower temperature. At a height of 1.05 $R_\odot$, the regions A, B, and C have $\log T_e$ of up to about 6.4, which is much higher than their vicinities. These are in agreement with the earlier findings \citep{Landi2008b}. The height variation of temperature in each region, on the other hand, is not clear in these maps. 

\begin{figure*}
\plotone{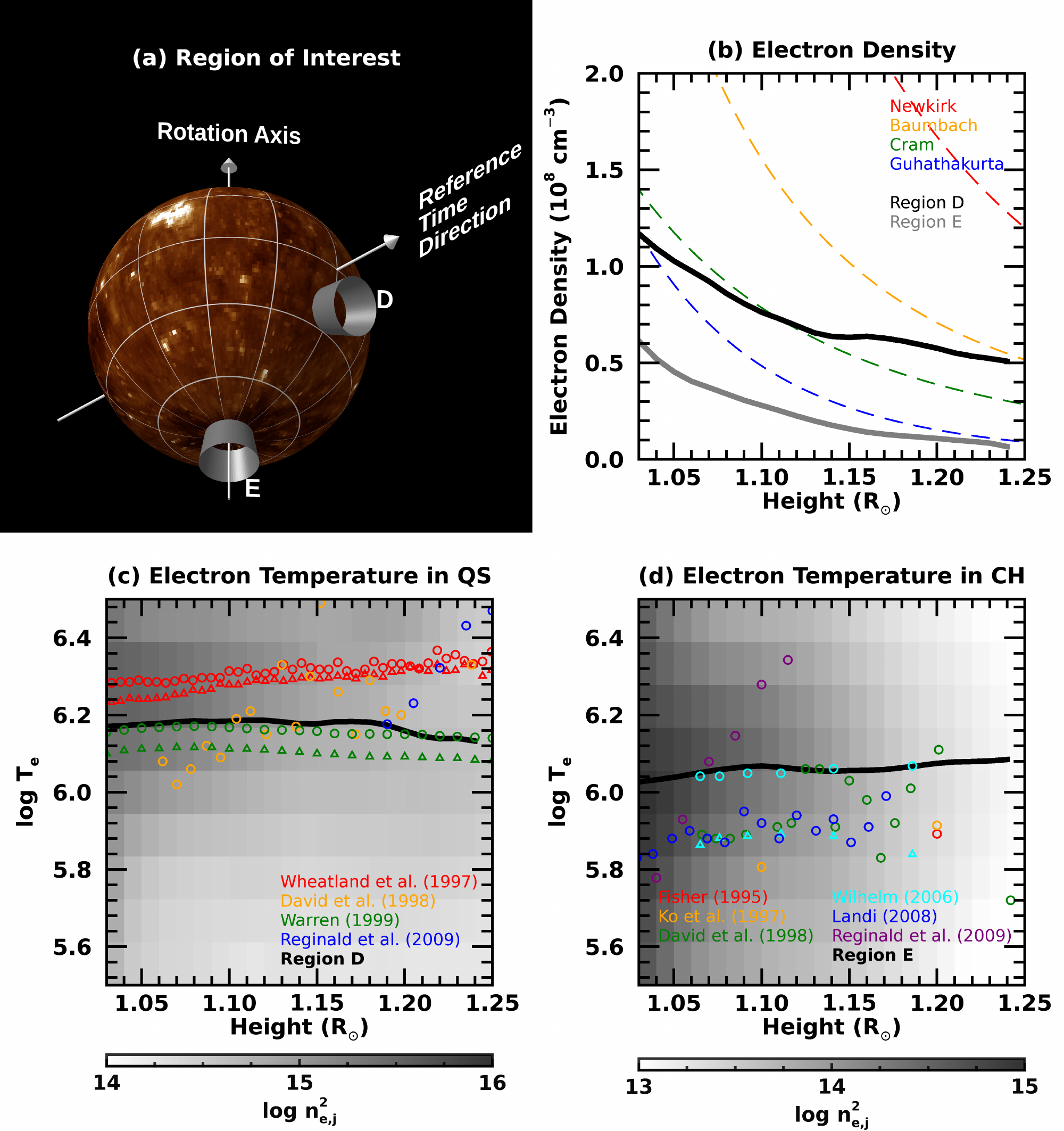}
\caption{(a) 3D view of the environment for the reconstructed cube. The solar surface is covered with the SDO/AIA 193 \AA\ Carrington map. Two gray truncated cones indicated by D and E are areas representing the quiet Sun region and the coronal hole, respectively. (b) Electron density profiles with height. Dashed lines with different colors represent various coronal density models. Black and gray solid lines indicate spatially averaged electron density profiles in regions D and E, respectively. (c) Electron temperature profiles with the height of quiet Sun regions. (d) Electron temperature profile with the height of coronal holes. Symbols with different colors present reported electron temperature profiles. The black solid line in both plots represents the mean electron temperature of each region. Gray maps in the background show the distribution of spatially averaged $n_{e, j}^2$ as a function of height and $\log T_e$. Note that different scale of $n_{e,j}^2$ distribution between (c) and (d).}
\label{fig9}
\end{figure*}

In order to perform more detailed analyses, we chose two specific regions (Figure \ref{fig9}a). Region D represents a quiet Sun region, and region E represents a coronal hole located in the south pole. Each of these regions has a volume of a truncated cone shape with an opening angle of 10 degrees. The positions of the central axis are (0\degr, -20\degr) and (0\degr, -90\degr), respectively, in heliographic coordinates. In both of the regions, electron density decreases monotonically with height (Figure \ref{fig9}b). The mean electron density profile of region D is comparable to the Cram model \citep{Cram1976} agreeing with \citet{VandeHulst1950} minimum equator model. The electron density profile in region E, which is $5 \times 10^7\, cm^{-3}$ lower than that of region D, is close to half of the Guhathakurta model \citep{Guhathakurta1999} for a polar coronal hole. The levels of $n_{e, j}^2$ distribution in Figure \ref{fig9}c and \ref{fig9}d also imply the difference of the electron density.

\begin{deluxetable*}{ccll}
\tablecaption{Previous studies for coronal electron temperature measurements \label{previous}}
\tablenum{1}
\tablewidth{0pt}
\tablehead{\colhead{Name} & \colhead{Instrument} & \colhead{Method}}
\startdata
\citet{Wheatland1997}\tablenotemark{a} & Yohkoh/SXT & Intensity ratio of Al 1265\AA\ to Al/Mg/Mn filter \\
\citet{David1998} & SOHO/CDS \& SUMER & Line ratio of O VI 1032\AA\ to 173\AA\   \\
\citet{Warren1999}\tablenotemark{b} & SOHO/TRACE & Intensity ratio of 171\AA\ to 195\AA\ image \\
\citet{Reginald2009}\tablenotemark{c} & ISCORE & Intensity ratio of 3850\AA\ to 4100\AA\ filter \\
\hline
\citet{Fisher1995} & \begin{tabular}[c]{@{}l@{}} Spartan WL coronagraph \\ \& Mk-III K-coronameter \end{tabular} & Radial electron density gradient\tablenotemark{d} \\
\citet{Ko1997} & Ulysses/SWICS & Various ionic ratios  \\
\citet{David1998} & SOHO/CDS \& SUMER & Line ratio of O VI 1032\AA\ to 173\AA\  \\	
\citet{Wilhelm2006}\tablenotemark{e} & SOHO/SUMER & Line ratio of Mg IX 706\AA\ to 750\AA\  \\
\citet{Landi2008a} & SOHO/SUMER & EM loci technique using various UV lines  \\
\citet{Reginald2009}\tablenotemark{c} & ISCORE & Intensity ratio of 3850\AA\ to 4100\AA\ filter  \\
\enddata
\tablenotetext{a}{Two different quiet regions}
\tablenotetext{b}{Used two different ionization balance calculation}
\tablenotetext{c}{Total eclipse observation}
\tablenotetext{d}{Scale height temperature}
\tablenotetext{e}{For plume(triangle) and lane(circle)}
\tablecomments{The upper and lower part indicate the quiet Sun and coronal hole studies, respectively.}
\end{deluxetable*}

We found that the mean electron temperature in region D varies little with height, with a value of \lgt\ = 6.17 (Figure \ref{fig9}c). The distribution of $n_{e, j}^2$ similarly indicates that most of the electrons have temperatures between $\log T = 6.1$ and 6.4.  Similarly, the mean electron temperature in region E also does not show temperature variation with height (Figure \ref{fig9}d). The mean value is \lgt\ = 6.06, which is about $3.4 \times 10^5$ K lower than region D. $n_{e, j}^2$ of region E is widely distributed with center of $\log T$ = 6.0. Both results are compatible with the previous studies for the electron temperature of quiet Sun and coronal hole (Table \ref{previous}). 

\subsection{Synthetic Coronagraph $pB$ Image}

\begin{figure*}
\plotone{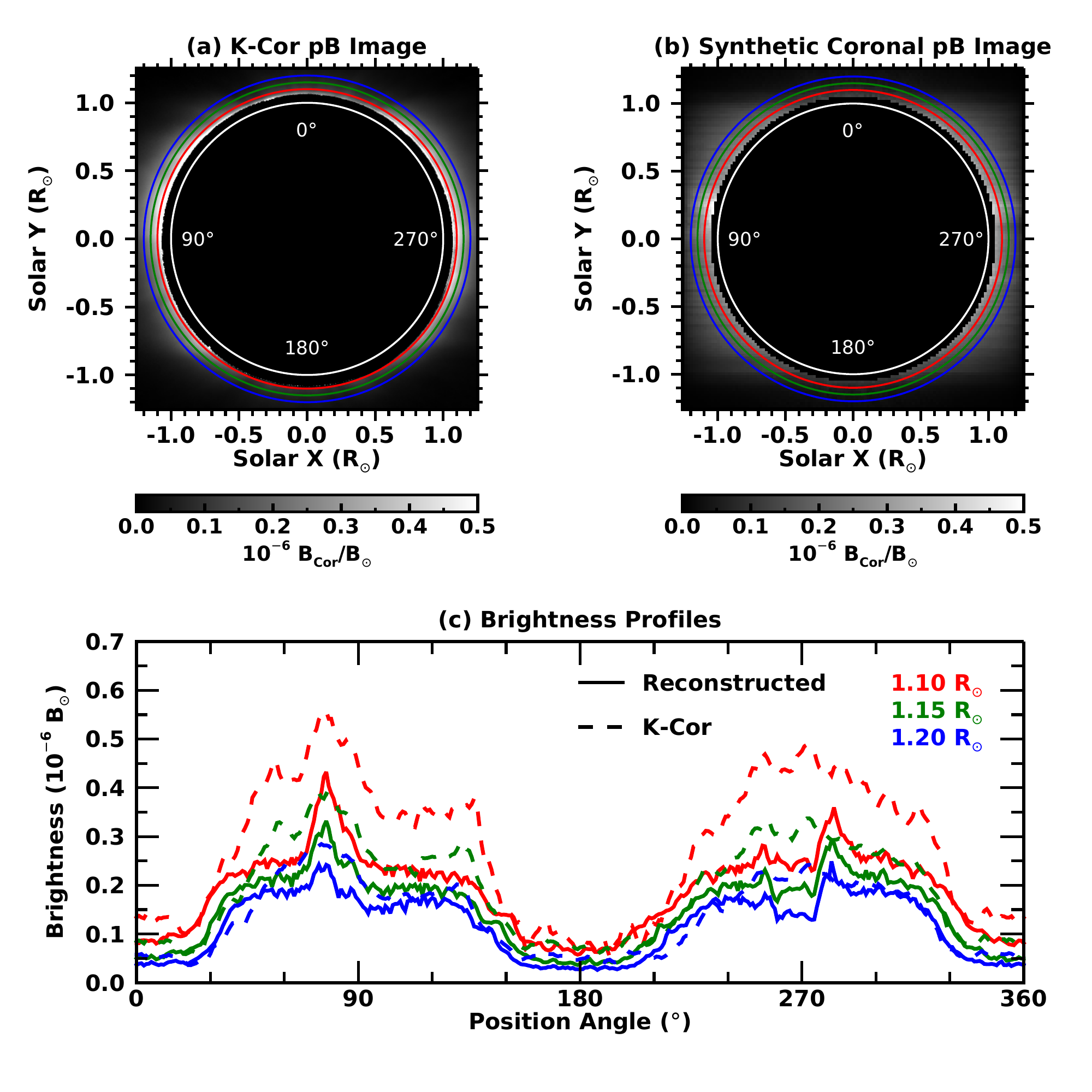}
\caption{(a) Observed $pB$ image from K-Cor observation. (b) Synthetic coronal $pB$ image. The position angles are marked on both images. (c) Brightness profile as a function of the position angle. Solid lines and dashed lines indicate data from the synthetic image and the observed image, respectively. Each color represents a distance from the solar disk center as indicated in (a) and (b). }
\label{fig10}
\end{figure*}

The synthetic coronal $pB$ image is overall similar to the real coronagraph $pB$ image (Figure \ref{fig10}a, \ref{fig10}b). The brightness variation as a function of position angle at different heights is shown in Figure \ref{fig10}c. In both the images, the brightness diminishes with height. It is also evident that the brightness in both images becomes low in the polar regions (0\degr\ and 180\degr) and high in the equatorial regions (90\degr\ and 270\degr). In the polar and higher regions, the reconstructed data are in good agreement with the real data, but, in the equatorial regions or at lower regions, the reconstructed data deviate from the real data. We suspect two possibilities for that reason. First, the error may arise from both AIA and K-Cor calibration processes (e.g. \citealt{Morgan2015}). Second, at the very lower height, especially in the equatorial region, the EUV emission may be optically thick \citep{Frazin2009}, which is against our assumption for the SRT.

Note that this result is calculated using the FBP algorithm with the EUV observation data, not the coronagraph observation data.  As Thomson scattering has scattering angle dependence and an occulter in coronagraph hides both frontside and backside simultaneously (areas 2 and 4 in Figure \ref{fig2}b), the coronagraph observation data is not suitable to use in the FBP algorithm. 

\section{Conclusion and Discussion} \label{sec:diss}

We have presented how the FBP algorithm can be applied to SRT. We tested the FBP algorithm with a modified sinogram using a test image, and applied it to SDO/AIA data. The FBP algorithm restores the relatively long-lived bright regions and the coronal holes seen in the real observations. The obtained physical quantities are compatible with the values of the accepted models and the previous observational studies. The synthetic coronagraph image well matches the observed K-Cor image. We emphasize that with the modified sinogram, the FBP algorithm can be conveniently used for the coronal studies. This simple algorithm has a significant advantage of efficient computation. The FBP algorithm with a modified sinogram allows us to readily probe into the solar corona. 

We assumed the parallel beam projection and the zero axial tilt angle. The error caused by parallel beam assumption increase merely 6\% (Figure \ref{fig3}). Thus we conclude that the parallel beam assumption is acceptable. In the case of the zero axial tilt angle assumption, on the other hand, the standard deviation of the difference volume increase about 3 times as tilt angle increase (Figure \ref{fig4}b). To ascertain the effect of the tilt angle on the SRT, we made another 193 \AA\ synthetic image through the same process as Section \ref{sec:res_1}. New data was observed in 2019 June when the tilt angle is close to zero. The result shows that standard deviation of the difference image is 30.7, which is a smaller but not much different value from the result of 2019 February data with about -7 degree tilt angle (See Figure \ref{fig6}d). This result demonstrates that the major source of the error is not the tilt angle but the temporal evolution of the solar corona. Therefore, we conclude that even though data observed in June or December guarantees better reconstruction quality, the difference in error would not be significant.  

The major limitation of the FBP algorithm is that it does not take into account the temporal evolution of the coronal structures. Since coronal structures change, the reconstructed data come to contain temporally averaged information and the resulting error. Thus, the error becomes large in the areas showing violent changes. 

The temporal evolution of coronal structures affects the obtained physical quantities. We have addressed the problem due to the negative values of $n_{e,j}^2$ in Section \ref{sec:data}. They may arise in principle from the incomplete reconstruction of the 2D Fourier domain, and we suppose that the temporal evolution of the solar corona is practically responsible for that error. Other SRT methods using the iterative reconstruction were similarly affected by that problem \citep{Frazin2002, Frazin2007, Frazin2009}, and handled it by constraining positive solution \citep{Frazin2000}. In our case, we cannot impose such positivity constraints, and hence we simply avoid it by replacing the negative $n_{e,j}^2$ voxels with zero value.

Consequently, limited objects in the solar corona can be investigated by using the FBP algorithm because of its static state assumption. Structures that are stable for more than two weeks, such as coronal holes, quiet Sun regions, and filaments, are reasonable subjects to study using the FBP algorithm. On the other hand, it is not possible to reconstruct transient events that change over timescales less than two weeks. If we can observe the solar corona at different angles using multi-spacecraft simultaneously, the problem related to the temporal evolution of the solar corona can be addressed \citep{Davila1994}.

 Further work is required to make better use of the DEM inversion method. The inferred physical quantities heavily rely on the DEM inversion method. They will vary with the choice of range for temperature, bin size of temperature, and Gaussian basis function of the inversion parameter \citep{Su2018}, or even different inversion methods \citep{Morgan2019a}. However, our FBP algorithm is totally independent of the specific DEM inversion method, and may be combined with any appropriate method. In principle, a better DEM inversion method will produce a better estimate of physical quantities with the help of our FBP algorithm.

\acknowledgments
We greatly appreciate the referee's constructive comments. We thank Mark Cheung for giving us a special lecture about DEM inversion method in KASI. This work was supported by the Korea Astronomy and Space Science Institute under the R\&D program (Project No. 2017-1-851-00 and Project No. 2020-1-850-07) supervised by the Ministry of Science and ICT. Courtesy of the Mauna Loa Solar Observatory, operated by the High Altitude Observatory, as part of the National Center for Atmospheric Research (NCAR). NCAR is supported by the National Science Foundation.

\end{document}